\documentclass[aps,prb,twocolumn,prX,showpacs]{revtex4}

\usepackage{boxedminipage}  
\usepackage{lscape}     
\usepackage{float}      
\usepackage[ansinew]{inputenc}  
\usepackage{epsfig}
\usepackage{subfigure}
\usepackage{graphicx}
\usepackage{amsmath}
\usepackage{amssymb}

\renewcommand{\d}[2]{\frac{#1}{#2}}
\newcommand{\pd}{\partial}

\begin{document}

\hyphenation{Brillouin}

\title{Magnetotransport in the insulating regime of Mn doped GaAs}

\author{Louis-Fran\c cois Arsenault}
\email{lfarsena@physique.usherbrooke.ca} \altaffiliation{Present address: Département
de Physique and RQMP, Université de Sherbrooke, Sherbrooke, Qc, Canada}
\affiliation{Department of Physics and Astronomy, Rutgers
University, Piscataway, NJ 08854}
\author{B. Movaghar}
\altaffiliation{Present address: Department of Electrical and Computer Engineering,
Northwestern University, Evanston, IL, USA}
\author{P. Desjardins}
\author{A. Yelon}

\affiliation{D\'epartement de G\'enie Physique and Regroupement
Qu\'eb\'ecois sur les Mat\'eriaux de Pointe (RQMP)\\ \'Ecole
Polytechnique de Montr\'eal, C.P. 6079, Succursale ``Centre-Ville'',
Montr\'eal (Qu\'ebec), H3C 3A7, Canada}

\date{\today}

\begin{abstract}
We consider transport in the insulating regime of GaMnAs. We
calculate the resistance, magnetoresitance and Hall effect, assuming
that the Fermi energy is in the region of localized states above the
valence band mobility edge. Both hopping and activated band
transport contributions are included. The anomalous Hall current
from band states is very different from the hopping Hall current and
has extrinsic (skew) and intrinsic (Luttinger) contributions. Comparison with experiment allows us to assess the
degree to which band and hopping contribution determine each of
the three transport coefficients in a particular temperature
range. There are strong indications that the insulating state
transport in GaMnAs is controlled primarily by extended state, band edge,
transport rather than by variable range hopping, as reported in the
literature.
\end{abstract}

\pacs{75.50.Pp, 75.47.-m, 72.20.Ee, 72.15.Rn}

\maketitle

\section{Introduction\label{sec:intro}} Magnetically doped
semiconductors constitute a very active field of research, for
good reasons. As magnets, they are expected to keep some of the
useful properties of the host system. As doped semiconductors, they
have their own usual applications now combined with new
functionality. For example, with Quantum Well (QW) multilayers and Quantum Dot (QD) Stranski-Krastanov growth technology\cite{Holub}, it is possible that spins are localized
in small clusters as in InMnAs\cite{Blatnner_Wessels} and in QD layers of MnGaAs\cite{Holub}. This could be used in building small magnets, memories
and spintronic switches. Therefore, localized state transport, magnetism of cluster-localized
states, and tunneling are very interesting and important phenomena. However, we must begin with a clear understanding of the bulk materials,
 whose electronic transport properties
in the insulating regime we investigate here. We demonstrate
that the observed properties cannot be explained solely by hopping
in homogeneous media. For this purpose, we need to develop the
hopping theory alongside the band theory and compare their
predictions. The approach in this paper has a wide range of
applications and is, we believe, useful to the wider magnetism and
device communities. The theory will enable us to treat resistance,
magnetoresistance, thermopower, normal and anomalous Hall effect in
the high resistance regime of DMS.\\
\\
Mn doped semiconductors\cite{Jungwirth1} are disordered alloys, so
one cannot expect sharp band edges, and indeed, one should observe
Anderson localized states above (for holes) the mobility edges. The
long range ferromagnetism is mediated by holes, which lower
their energy each time they interact with a Manganese spin of
opposite spin. The mobile valence band $p$-holes interact with the
localized moments via the antiferromagnetic exchange coupling
\begin{equation}\label{double_echange}
H_{pd} = \d{J_{pd}}{2}\sum_{m \in
\{N_S\}}\sum_{s,s'}c_{ms}^{\dagger}\boldsymbol{\sigma}_{ss'}\cdot\textbf{S}_mc_{ms'},
\end{equation}
where $J_{pd}$ is the hole-Mn coupling, $\boldsymbol{\sigma}_{ss'}$ is a vector containing the Pauli's
matrices i.e. $[\sigma^x, \sigma^y, \sigma^z]$. The indices, $s$ and
$s'$ indicate which terms of the 2x2 matrix we are considering.
Finally, $\textbf{S}_m$ is the Mn spin operator at site m and the
$c_{ms}^{\dagger}$, $c_{ms}$ are creation and annihilation operators
for a carrier of spin $s$ at site $m$. This Hamiltonian, applied to
GaMnAs, has been discussed in many papers and we refer the reader to
the original literature (Ref.~\onlinecite{Jungwirth1} and references
therein). The transport and magnetism was recently subjected to a
coherent potential approximation (CPA) treatment\cite{Arsenault1}.
The CPA is a mean field theory which can handle delocalized states up
to the short mean free path or diffusive limit, but it does not
reproduce localization. Localized states and hopping transport have
to be treated separately, as shown below.\\
\\
Near the band edges, the eigenstates of one particle systems are localized for energies up to the mobility edges
$\varepsilon_c$\cite{Mott_Davis}. The localization starts when the
density of states is below a critical value determined by the
diffusivity and localization length\cite{Movaghar}. The nature of
the mobility edges in disordered magnets is expected to be more
complex than that of nonmagnetic alloys, but many of the usual
features should remain valid. In a magnet such as GaMnAs, we now
have two spin-split bands, and two mobility edges. Though
conductivity, magnetoresistance and Hall coefficient have been
measured in the insulating limit\cite{Van_Esch1,Allen}, the dependence
of the mobility edges on magnetism has not been investigated.
Neither is it clear from the present understanding of the data
whether all transport coefficients are truly in the hopping limit,
or whether there is a substantial band edge delocalized contribution
which looks like hopping. One usually says that transport is in the
hopping limit when the measured conductivity scales with temperature
as $\text{exp}\left[ -\left(\d{T_0}{T}\right)^{\d{1}{4}} \right]$.
Here, $T_0$ is the renormalized Mott temperature $T_0 = 24\times
2.7\alpha^3/(\pi\rho(\varepsilon_f)k_B) $ with $\alpha$ denoting the
inverse localization length\cite{Mott_Davis}, $k_B$ the Boltzmann
constant, and $\rho(\varepsilon_f)$ is the density of states at the
Fermi level and 2.7 the percolation factor. But hopping transport
with Coulomb correlation gaps often scale as $\text{exp}\left[
-\left(\d{T_0}{T}\right)^{\d{1}{2}} \right]$, and there are many
situations in which materials exhibit $\text{exp}\left[
-\left(\d{T_0}{T}\right)^{\mu} \right] \rightarrow 0 < \mu < 1$
laws, which are obviously not due to variable range hopping (VRH)
between localized states (see Ref.~\onlinecite{Anglada}).
Emin\cite{Emin}, for example, has shown that small polarons will
also produce such behavior, for reasons not related to VRH.
Indeed, one can say that almost all disordered insulators exhibit
such a weaker-than-exponential law. Often it is due to granularity
and Coulomb correlations\cite{Anglada,Dunford}. Furthermore, it is
often not easy to distinguish between $\mu \sim 1/3, 1/4 \ldots$etc
because the data are collected over a limited range. So one has to
be cautious and examine other possible scenarios. This is what we
propose to do here.\\
\\
We do this by presenting a complete theory of magneto-transport
which is useful for other materials as well. But, as an application of
the formalism developed here, we concentrate our analysis on Mn doped GaAs. On the basis of
experimental data for magnetoresistance and thermopower, we question
the assumption of Mott-hopping in inslating GaMnAs and we offer an alternative explanation.
Concerning the Hall effect, we recall that Allen \emph{et al}.\cite{Allen} have shown that the Anomalous Hall
Effect (AHE) in the insulating regime of GaMnAs is not just
anomalous in the usual sense because it scales with magnetism, but
also has an anomalous sign. An explanation of this anomaly is
apparently contained in the hopping Hall effect model of Burkov and
Balents\cite{Burkov_Balents}. Here, we propose an
alternative theory which includes both hopping and band edge
conduction.\\
\\
This paper is structured as follows: we first
present the basic definition of the Hall effect in magnets. The
processes that explain transport in the insulating limit are then
discussed. All the main contributions to the Hall effect are then
introduced, including hopping, intrinsic and extrinsic, and the
behavior near the mobility edge. We then discuss the
magnetoresistance and the different contributions to it are
introduced. Following that, all the contributions to the transport
propertires are summarized and discussed in relation with
experiments. Finally, we present a discussion of our results. We
argue that transport in the insulating regime of the magnetic
semiconductors of the GaMnAs type above 10K, is mainly one of
delocalized band edge conduction.
\section{Hall effect}
The experimentally measured Hall
coefficient $R_H$ is often written as
\begin{equation}\label{eq:RHall_1}
\begin{split}
R_H &= R_N + R_A\\
R_A &= \left(a\rho_{xx}+b\rho_{xx}^2\right)M_z/B_z^{ext},
\end{split}
\end{equation}
where $a$,$b$ are constants and $\rho_{xx}$ is the resistivity.
$B_z^{ext}$ is the external magnetic induction, defined as
$\mu_0H_z^{ext}$, where $H_z^{ext}$ is the external magnetic field.
The first term $R_N$ is the normal Hall coefficient and scales with
resistivity in the usual way and the second $R_A$ is the anomalous
term which, in general, can have two components, one linear and the
other quadratic with resistivity\cite{Jungwirth1} and is
proportional to the magnetization $M_z$. The general relation for
the Hall coefficient is\cite{Arsenault2}
\begin{equation}\label{RHall_rigor}
R_H =
\d{\rho_{yx}}{B_z^{ext}},
\end{equation}
with
\begin{equation}\label{rho_yx}
    \rho_{yx} = \d{ -\text{Re}\left\{\sigma_{xy}\right\}  }{ \text{Re}\left\{\sigma_{xx}\right\}\text{Re}\left\{\sigma_{yy}\right\} - \text{Re}\left\{\sigma_{yx}\right\}\text{Re}\left\{\sigma_{xy}\right\}},
\end{equation}
where $\sigma_{xy}$ and $\sigma_{xx}$ denote the transverse and
normal conductivity respectively. We may suppose that, as there is
no voltage applied in the $y$ direction $\sigma_{yx}\sigma_{xy} \ll \sigma_{xx}\sigma_{yy}$.
Also, if we consider an isotropic system in the xy plane $\sigma_{xx}=\sigma_{yy}$ and thus we may write the Hall coefficient
as\cite{Arsenault1,Movaghar_Cochrane1}
\begin{equation}\label{RHall_1}
R_H =
\d{\text{Re}\left\{\sigma_{xy}\left(B_z\right)\right\}}{B_z^{ext}\left[
\text{Re}\left\{\sigma_{xx}\left(B_z\right)\right\} \right]^2}.
\end{equation}
Karplus and Luttinger\cite{Karplus_Luttinger} pointed out that the
$B$-field involves the magnetic moment of the material via the
internal magnetization $M_z$
\begin{equation}\label{mag_field}
B_z = \mu_0\left[ H_z^{ext} + (1-N)M_z \right] \equiv B_z^{ext} +
\mu_0(1-N)M_z,
\end{equation}
where $N$ is the demagnetizing factor. Thus, the magnetization term
is implicit in the normal contribution as a shift in the magnetic
field. However this form is not normally sufficient to explain the
much larger magnetization contribution observed in
ferromagnets\cite{Karplus_Luttinger}. Karplus and Luttinger then
developed the first theory of the intrinsic
AHE\cite{Karplus_Luttinger}. Throughout, we shall, for simplicity,
suppose a thin film with the magnetic field perpendicular to the
plane and thus, we take $N = 1$, unless otherwise mentioned. Let us
now consider the problem of conduction.
\section{Hall effect and conductivity in the insulating limit}
When the Fermi level is at the band edge, Anderson localization sets
in, and Fermi level transport cannot be described using only extended
states. The very low temperature transport process is by hopping, as
shown by Allen \emph{et al}.\cite{Allen} for GaMnAs. These authors
have also measured the Hall effect and, as already mentioned above,
observed that there is a sign anomaly for the AHE in this regime.
They argue that the transport is by hopping from localized level to
localized level and not by thermally assisted hole ionization into
the valence band. The magnetoresistance in the insulating regime has
been measured by Van Esch \emph{et al}.\cite{Van_Esch1}. These
authors constructed a model to explain the very large
magnetoresistance they observed, which we will discuss later. But no
one seems to have yet made a model which consistently describes
resistance, magnetoresistance, and Hall effect, in the insulating
regime of GaMnAs.\\
\\
The ordinary phonon assisted hopping Hall effect has already been treated
rigorously and in great detail\cite{Movaghar1,Movaghar2,Movaghar3}. The extraordinary
contribution reported recently by Allen \emph{et al}.\cite{Allen} in
Mn doped GaAs, can be modeled as follows. We consider $p$-holes in
the valence band, spin-orbit coupled as in the models of Baldareshi
and Lipari\cite{Baldareshi_Lipari} and Fiete \emph{et
al}.\cite{Fiete}. The holes are bound to a negatively charged, spin
5/2, Mn ion. Fiete \emph{et al}.\cite{Fiete} also
include the AF exchange coupling of the hole spin to the $d$-spin.\\
\\
The Hamiltonian describing the dynamics of holes localized around
the Mn sites and which incorporates the spin-orbit coupling to first
order is given by
\begin{equation}\label{Hamiltonian}
\begin{split}
    H &= \sum_{i,s}\varepsilon_{is}c_{is}^{\dagger}c_{is} +
    \sum_{i,j,s}t_{ij}(B_z+B_{so})c_{is}^{\dagger}c_{js}\\ &-
    \sum_{i,s,s'}\lambda_i\langle
    i,s|\textbf{l}_i\cdot\boldsymbol{\sigma}|i,s'\rangle
    c_{is}^{\dagger}c_{is'}\\
    &- \sum_i\mu_B\left( l_{i,z}+g\sigma_{i,z} \right)\left( B_z + h_{z,spin}
    \right)c_{is}^{\dagger}c_{is}.
\end{split}
\end{equation}
In Eq.~\eqref{Hamiltonian}, the summation indices include the band
and site indices, i.e. $j = {j,\gamma}$ where $\gamma$ denotes the
orbital, and we again assume $N = 1$. The $\varepsilon_{is}$ are
diagonal site energies. Now consider, the second term, where
$t_{ij}(B_z+B_{so})$ is the site-to-site transfer term. The magnetic
field is incorporated in the Peierls phase. We recall that in the
presence of the magnetic field we have the usual Peierls phase
factor
\begin{equation}\label{Peierls}
    t_{mn}=t_{mn}^0\text{e}^{-\d{ie}{2\hbar}\textbf{B}^{ext}\cdot(\textbf{R}_n\times\textbf{R}_m)}.
\end{equation}
The non-local spin-orbit terms are not specifically included in
Eq.~\eqref{Hamiltonian} but can be reduced, in first order, to a
shift of the total magnetic field by an amount $B_{so}$, that is
included in $t_{ij}$ and which we consider in detail below. The
atomic spin-orbit coupling $\lambda_i$, in the third term in
Eq.~\eqref{Hamiltonian}, mixes angular momentum \textbf{l} and spin
bands on the same atom. Assuming, for the sake of argument, that we
have orbital states which are $p$-states with $l_z = \{1,0,-1\}$ or
combinations thereof, then the coupling will flip, for example,
$|1,-1/2\rangle \rightarrow |-1,1/2\rangle$ and back again. The
z-component of the atomic spin-orbit term ( $-\lambda
l_{i,z}\sigma_z$ ) can be included in the sum of the Zeeman and
Curie Weiss spin splitting energy for the carriers, where $\mu_B$ is
the Bohr magneton. This sum is denoted by $h_{z,spin}$ in the fourth
term of Eq.~\eqref{Hamiltonian}.\\
\\
The spin-orbit Hamiltonian in tight-binding has been examined by
Pareek and Bruno\cite{Pareek_Bruno} for disorder induced coupling,
by Movaghar and Cochrane\cite{Movaghar_Cochrane1} for longitudinal
and Hall transport and in CPA by
Arsenault \emph{et al}.\cite{Arsenault1,Arsenault2}. In weak spin-orbit coupling,
the effect can be incorporated into a two-site overlap phase similar
to the Peierls phase given by
Eq.~\eqref{Peierls}\cite{Damker,Movaghar_Cochrane1,Arsenault1}. In
the hopping limit, with no external field, the electric field at a
site i (multibands are implicit if necessary) is given by the sum of
the fields emanating from all neighbors
\begin{equation}\label{Eso}
\textbf{E}_{so,i} = \left\langle
i\left|\sum_n\d{eZ_n(\textbf{r}-\textbf{R}_n)(1-p_n)}{4\pi\varepsilon\varepsilon_0|\textbf{r}-\textbf{R}_n|^3}\right|i\right\rangle
.
\end{equation}
In Eq.~\eqref{Eso}, $\varepsilon$ is the dielectric constant which
screens the acceptor site potential, $Z_n$ is the effective charge,
and the factor $(1-p_n)$ is the probability that the site $n$ is
empty and therefore charged. In the i to j jump, the third closest
site in the pathway will contribute an electric field which produces
a net magnetic field. So the sum in Eq.~\eqref{Eso} can be
approximated to include the best neighbor to i (see Refs.~\onlinecite{Movaghar}, \onlinecite{Movaghar2}, \onlinecite{Gruenewald} and refs therein, for the definition of best neighbor). We can now include
this spin-orbit phase in the magnetic phase by adding to $B^{ext}$
in Eq.~\eqref{Peierls} the simplified effective spin-orbit field
\begin{equation}\label{Bso}
\textbf{B}_{so}(i\rightarrow j) = \d{\hbar}{2mc^2} \left\langle
i\left| \sum_{n,n\neq
i,j}\d{Z_n(1-p_n)}{4\pi\varepsilon\varepsilon_0|\textbf{r}-\textbf{R}_n|^3}
\right|i\right\rangle\boldsymbol{\sigma},
\end{equation}
where the nearest neighbor sum excludes the start site i and end
site j. A more complete derivation for the effective spin-orbit
field is presented, for the case with no screening (no $\varepsilon$
in the equation) and where all potentials contribute (no $1-p_n$ in
the equation), in Ref.~\onlinecite{Arsenault1}. Note that the integral $\langle i |\d{1}{|\textbf{r} - \textbf{R}_n|^3}|i\rangle$is
strictly speaking not convergent. But, in practice, the orbit radius is never allowed to be smaller
than the effective atomic orbit of the valence state so that the cubic singularity does not occur.\\
\\
Before determining the various contributions to the Hall conductivity, let us examine the
effective masses which may appear in the expressions which follow, so that the meaning is clear. There
 are three effective masses in the problem:
\begin{enumerate}
  \item
  the Kane-Luttinger mass ($m^*$) which controls the kinetic energy in the starting Hamiltonian (effective mass Hamiltonian);
  \item
  the spin-orbit mass which is equal to the bare electron mass
  \item
  an effective mass which is a result of the sum rule, as discussed in Refs.~\onlinecite{Datta} and \onlinecite{Arsenault_Movaghar}, and which can become the diffusivity in strong disorder but is the Kane-Luttinger mass in weak disorder.
\end{enumerate}
\subsection{Hopping contribution}
The normal Hall term arising from the external $B=B_z$ term has been
derived by the hopping interference method in the pure diffusive
limit, and in the VRH limit\cite{Movaghar2,McInness,Burkov_Balents}
and applied to the case of phonon assisted hopping and percolation
in a one band model. Thus, we may add the spin-orbit magnetic field
to the external field and carry out the configurational average as
before. So we replace $B_z$ by
\begin{equation}\label{B_tot+so}
    B_z^{ext} + \mu_0(1-N)M_z + B_{so},
\end{equation}
where
\begin{equation}
\begin{split}
    B_{so} &= \left\langle i\left| \d{\hbar}{2mc^2}\sum_{n,n\neq
i,j}\d{Z_n(1-p_n)}{4\pi\varepsilon\varepsilon_0|\textbf{r}-\textbf{R}_n|^3}\right|
i\right\rangle\langle\sigma_z\rangle \\
&\equiv \lambda_{hop}\langle\sigma_z\rangle,
\end{split}
\end{equation}
where $\langle\sigma_z\rangle$ is the polarization of the holes and
$M_z$ the magnetization of the sample. The analysis  can now be done
as previously. The high density hopping system, for example, has
been treated analytically in Ref.~\onlinecite{Movaghar2}, and the
phonon assisted percolation regime in Refs.~\onlinecite{Movaghar1}, \onlinecite{Movaghar3}, and ~\onlinecite{Gruenewald}.\\
\\
The second term in Eq.~\eqref{B_tot+so} depends upon the geometry of
the sample and is zero in thin films under normal
fields\cite{Karplus_Luttinger}. The sum in the third term runs over
the third neighbor in the triad and is configurationally averaged to
the optimum second nearest neighbor distance (as shown in
Ref.~\onlinecite{Movaghar1}), which is $|\textbf{R}_i-\textbf{R}_n|
= \left\langle R_3(T) \right\rangle \propto
\left(\d{T_0}{T}\right)^{\d{1}{4}}$. Here, $T_0$ is the
renormalized Mott temperature, defined previously.\\
\\
Replacing Eq.~\eqref{B_tot+so} in the expression of
Ref.~\onlinecite{Gruenewald} allows us to calculate the Hall
mobility (sign is negative for electrons and positive for holes) and
conductivity as was done before\cite{Gruenewald}. Assuming a
constant density of states, we obtain
\begin{equation}\label{mob_xy}
    \mu_{xy} =
    \d{\text{Re}\{\sigma_{xy}\}}{B_z^{ext}\text{Re}\{\sigma_{xx}\}} =
    \text{e}^{-\d{3}{8}\left(\d{T_0}{T}\right)^{\d{1}{4}}}\left[ 1 +
    \d{\langle\sigma_z\rangle}{B_z^{ext}}\lambda_{hop}
    \right],
\end{equation}
\begin{equation}\label{conduc_hop}
    \text{Re}\{\sigma_{xx}\} =
    \sigma_0\text{e}^{-\left(\d{T_0}{T}\right)^{\d{1}{4}}} =
    \d{1}{\text{Re}\{\rho_{xx}\}},
\end{equation}
and
\begin{equation}\label{rho_xy_div_B}
    \d{\rho_{xy}}{B_z^{ext}} =
    \d{\mu_{xy}}{\text{Re}\{\sigma_{xx}\}} = \text{e}^{\d{5}{8}\left(\d{T_0}{T}\right)^{\d{1}{4}}}\left[ 1 +
    \d{\langle\sigma_z\rangle}{B_z^{ext}}\lambda_{hop}
    \right].
\end{equation}
Here $\rho_{xy}$ is the Hall resistivity, $\sigma_0$ the
conductivity prefactor and we have used $N=1$. Note that for the
remainder of the paper we will no longer distinguish the real part
of the conductivity from total conductivity, since they are the same
at zero frequency.\\
\\
The factor 3/8 in Eq.~\eqref{mob_xy} is a consequence of averaging
over the third site in the triad $\langle 2\alpha R_3 \rangle \sim
\d{13}{8}\left(\d{T_0}{T}\right)^{\d{1}{4}}$. When the density of
states is quadratic in energy, rather than constant, we replace
$\left(\d{T_0}{T}\right)^{\d{1}{4}}$ by
$\left(\d{T_0}{T}\right)^{\d{1}{2}}$. This is also true for the
Coulomb gap hopping
problem\cite{Shlovskii_Efros,Gruenewald,Movaghar1,Movaghar_Schirmacher}.\\
\\
Allen \emph{et al}.\cite{Allen} observed an experimental
$\text{exp}\left[ -\left(\d{T_0}{T}\right)^{\d{1}{2}} \right]$ dependence of conductivity. This is characteristic
of a Coulomb gap or hopping in a quadratic density of states. They
also found a near $\rho_{xx}^{\d{1}{2}}$ behavior of the Hall
resistivity. This is also reproduced by the present theory as shown
in Eq.~\eqref{rho_xy_div_B}. Finally, they found normal and anomalous
processes which have opposite signs. The sign anomaly can be
explained by the sign of $\langle\sigma_z\rangle$ which can be
opposite to the overall
magnetization since it refers only to carriers at the Fermi level.\\
\\
The maximum strength of the AHE predicted by Eqs.~\eqref{mob_xy},
~\eqref{conduc_hop} and ~\eqref{rho_xy_div_B} is determined by the
magnitude $\lambda_{hop}$. An estimate which includes the factor
1000 reduction due to the cube of the average distance to the
$3^{rd}$ neighbor, gives $\sim 3\times 10^3$ Gauss. This spin-orbit
term is somewhat too small at $T=25$K  when the result is compared
to the experimental data of Ref.~\onlinecite{Allen}. The data
suggest that the AHE is more important than the Normal Hall effect,
even at an external field of $10$T. The measured order of magnitude
is close to the theory estimate only in the truly very low
temperature hopping regime where AHE $\sim$ NHE i.e when $T\sim 10$K
or less. The ratio of AHE to NHE is always roughly 5 or 6 when we
look at the data in the \emph{metallic regime}. It seems therefore
that though the simple hopping model may be right for the truly low
temperature hopping regime $T<10$K, at higher temperatures the
spin-orbit electrical fields which are needed to produce the
sideways force are larger than those predicted by
hopping theory.\\
\\
This is possibly because, in this regime, the Hall charge transport
is, in fact, already activated to above the mobility edge and is not
really hopping-like. We shall examine this possibility  in the next
section. This view is also supported by
magnetoresistance\cite{Van_Esch1} and thermopower
data\cite{Osinniy}(although Osinniy \emph{et al.}\cite{Osinniy} say
otherwise). Thus, the picture of the third site in the triad
providing the spin-orbit field is attractive, but the effect
produced by this mechanism is probably not large enough to explain
the quantitative values. States excited above the mobility edge
undergo skew scattering and intrinsic Hall scattering and feel the
full effect of the lattice-induced spin-orbit enhancement, as discussed
by Chazalviel\cite{Chazalviel}, Berger\cite{Berger},
Fivaz\cite{Fivaz} and Engel \emph{et al}.\cite{Engel}. The previous
work of De Andrada \emph{et al}.\cite{DeAndrada}, for example, gives
a maximum enhancement factor of the bare coupling of the form
\begin{equation}\label{enhan_so} \gamma_{enh} = \left( \d{m}{m^*}
\right)\d{mc^2}{E_g}\d{\Delta\left( 2E_g + \Delta \right)}{\left(
E_g + 3\Delta \right)\left( 3E_g + \Delta \right)},
\end{equation}
where $\Delta$ is the Kane spin-orbit coupling energy, $E_g$ is the
energy gap of the semiconductor and $m^*$ is the effective mass. One
can see that in GaAs, the enhancement of the spin-orbit coupling is
considerable. Thus, the pathways above the mobility edge, even if
not necessarily always dominant for the ordinary transport, can be
expected to be dominant for the AHE. It is difficult to see where
such an enhancement would come from in hopping conduction. Including
the many orbital bands will not strongly affect the results in the
hopping limit. The marginal but important contributions from the
hopping pathways forces us to now look for other possible
explanations, or other contributions to the hopping channel
specially at higher temperatures.
\subsection{Intrinsic Luttinger contributions from the region above the mobility edge}
Consider the contributions to transport coming from excited states
below (for holes) the mobility edge. For delocalized states, there are skew
scattering Hall contributions due to impurity potentials. This has
been treated in Refs.~\onlinecite{Engel} and
~\onlinecite{Ballentine}. There is also an intrinsic Luttinger
contribution which can be written, in the language of
Chazalviel\cite{Chazalviel}, and as rederived  to first order in
spin orbit coupling by Arsenault and
Movaghar\cite{Arsenault_Movaghar}. We write the transverse
conductivity as
\begin{equation}\label{trans_conduc}
    \sigma_{xy}^i = \d{e^2}{\Omega}\d{1}{e}\sum_{\alpha}\left(-\d{\pd
    f(\varepsilon_{\alpha})}{\pd\varepsilon_{\alpha}}\right)\mu_{B}\sigma_z^{\alpha}\Delta g_{\alpha}^{zz},
\end{equation}
where
\begin{equation}\label{def_sigz}
    \sigma_z^{\alpha} = \d{n_{\alpha\uparrow} - n_{\alpha\downarrow}}{n_{\alpha\uparrow} + n_{\alpha\downarrow}},
\end{equation}
$e = -|e|$ for electrons and $e = |e|$ for holes. Finally, the $zz$ component of the effective $g$-shift tensor is
\begin{equation}\label{del_g}
\begin{split}
    &\Delta g_{\alpha}^{zz} =\\ &\sum_{\beta}\left\langle \alpha
    \left|\d{\hbar}{4m^2c^2}\left( \nabla V(\textbf{r})\times\textbf{p}
    \right)_z\right|\beta\right\rangle\d{1}{\varepsilon_{\beta} -
    \varepsilon_{\alpha}}\left\langle\beta\left|L_z\right|\alpha\right\rangle.
\end{split}
\end{equation}
In the preceding equations, $\mu_{B} = \d{e\hbar}{m}$,
$f(\varepsilon_{\alpha})$ is the Fermi function, $n_{\alpha s}$ is the number of electrons in the state $\alpha$ with spin $s$ and $\Delta g_{\alpha}^{zz}$ is the $zz$ component of the $g$-shift tensor. This component
can be related to the result first derived by Karplus and Luttinger\cite{Karplus_Luttinger} by writing
approximately $\Delta g \sim \lambda/\Delta_{gap}$ where
$\Delta_{gap}$ is a (Bloch) subband gap and $\lambda$ the spin-orbit
coupling strength. The $g$-shift is used here as in the language of
Chazalviel\cite{Chazalviel}, and includes the Bloch band
enhancement. Note that the intrinsic contribution given by
Eq.~\eqref{trans_conduc} apparently does not scale with the
conductivity. But this is only true in a highly ordered medium in
which the scattering linewidth is smaller than the intersubband
energies. In this limit, it only appears as a lifetime term in a
basis of well defined Bloch states. In a strongly disordered system,
there are no well defined Kohn-Luttinger bands as such, and the
$g$-shift will contain the relaxation time term in the denominator
as well. When the linewidth is larger than the dominant excitation
energies in the sum, then the relaxation time will enter the
$g$-shift and the Luttinger term will also scale with resistance.
\subsection{Magnetic field dependent resistance and Hall effect contributions from the region above the mobility edge}
We can now give a  more complete formula which includes the hopping
channel and the mobility edge contributions and which allows us to
predict the magnetoresistance as well. From standard mobility edge
transport theory\cite{Mott_Davis,Khmelnitszkii}, and including the
normal Hall effect, the skew scattering
contribution\cite{Ballentine,Engel} and the intrinsic
Karplus-Luttinger term, we have
\begin{widetext}
\begin{equation}\label{3contri_trans_conduc} \sigma_{xy} =
e^2\int_{-\infty}^{\varepsilon_c}d\varepsilon\left(-\d{\pd
f_h(\varepsilon)}{\pd\varepsilon}\right)\rho(\varepsilon)D_{diff}(\varepsilon)\left(
\d{eB_z}{m(\varepsilon)}\tau (\varepsilon) +\langle \sigma^z
\rangle\d{\tau (\varepsilon)}{\tau_s(\varepsilon)} \right) +
\sigma_{xy,hopping}(T) + \sigma_{xy}^i,
\end{equation}
\end{widetext}
where the so called intrinsic AHE is
\begin{equation}\label{int_contri}
    \sigma_{xy}^i = e^2\int_{-\infty}^{\varepsilon_c}d\varepsilon\left(-\d{\pd
f_h(\varepsilon)}{\pd\varepsilon}\right)\rho(\varepsilon)\d{\hbar}{m^*}\Delta
g(\varepsilon)\sigma^z(\varepsilon)
\end{equation}
and
\begin{equation}\label{2contri_conduc}
\sigma_{xx} =
e^2\int_{-\infty}^{\varepsilon_c}d\varepsilon\left(-\d{\pd
f_h(\varepsilon)}{\pd\varepsilon}\right)\rho(\varepsilon)D_{diff}(\varepsilon)
+ \sigma_{hopping}(T).
\end{equation}
The second part of the first term in
Eq.~\eqref{3contri_trans_conduc} is the skew scattering
contribution with $\d{1}{\tau_s}$ denoting the skew scattering rate. Here, $f_h(\varepsilon)$ is the hole distribution. In
Eq.~\eqref{3contri_trans_conduc} and Eq.~\eqref{2contri_conduc},
$D_{diff}(\varepsilon)$ is the band diffusivity at energy
$\varepsilon$, and by localization theory\cite{Khmelnitszkii}, is of
the form
\begin{equation}\label{Diffusitivity}
    D_{diff}(\varepsilon) = D_0\left|\d{\varepsilon_c-\varepsilon}{\varepsilon_c}\right|
\end{equation}
near the mobility edge. Note that the Luttinger term
contains the quantum diffusivity $\hbar/m^*$ and not the actual diffusivity $D_{diff}$. In Eq.~\eqref{Diffusitivity} $D_0 \sim
1$cm$^2$/s, $\rho(\varepsilon)$ is the density of states for which
one normally adopts an exponential form:
\begin{equation}\label{dos_exp}
    \rho(\varepsilon) = \rho_0\text{e}^{ -\d{\varepsilon}{\varepsilon_0}
    },
\end{equation}
with $0 < -\varepsilon < -\varepsilon_c$, where $\rho_0$ is the
value at the Fermi level and $\varepsilon_0$ measures the
exponential steepness. The spin-orbit coupling is contained in the
$g$-shift factor. If the intersubband gaps $\Delta_{gap}$ is
definable, the $g$-shift is $\Delta g_{\alpha} \sim
\d{\lambda}{\Delta_{gap}}$ as in Karplus and
Luttinger\cite{Karplus_Luttinger}. In highly disordered alloys,
there are no well defined subband gaps, as assumed by Karplus and
Luttinger\cite{Karplus_Luttinger} and Jungwirth \emph{et
al}.\cite{Jungwirth1}. Then, one has to evaluate the $g$-shift more
rigorously. The $g$-value of the Mn spin + hole complex ($\sim$1.98)
has been measured by ferromagnetic resonance and is in the range
(-0.05 to -0.2), \emph{hole like} ($\Delta g < 0$), and increases with Mn-hole doping
(see the review by Liu and Furdyna\cite{Liu}). Note that the sign
anomaly can again, as in hopping, be due to the fact that the holes
at the Fermi level are polarized opposite to the bulk magnetism.\\
\\
The hole Fermi function $f_h(\varepsilon)$ is, to a good
approximation, a Boltzmann function when $\left|
\varepsilon_f-\varepsilon_c \right| \gg k_BT$. The mobility edge for
holes is denoted by $\varepsilon_c$, and the energy dependence of
$\rho(\varepsilon)D_{diff}(\varepsilon)$ dominates the longitudinal
conductivity. The anomalous skew scattering Hall contribution will
be dominated by the spin-orbit band term via
$\d{1}{\tau_s(\varepsilon)}$ in
Eq.~\eqref{3contri_trans_conduc}\cite{Ballentine,Engel}. The skew
scattering contribution is not negligible because of the very large
spin-orbit coupling enhancement ($10^4$), implicit in
Eq.~\eqref{enhan_so}, when applied to GaMnAs. It is useful to recall
that Engel \emph{et al}.\cite{Engel} have estimated the skew factor
to be $\d{\tau}{\tau_s} \sim \d{1}{900}$ for Si doped GaAs. Thus,
when comparing $\d{\tau}{\tau_s}\langle \sigma^z \rangle$ with
$\omega_c\tau$, and noting that $\tau \sim
\d{Dm^*}{\left|\varepsilon_c-\varepsilon\right|}$, $\varepsilon <
\varepsilon_c$ (for holes), we see that the skew scattering band
contribution to the AHE is comparable to the ordinary Hall effect
even near band edges. The Luttinger term (Eq.~\eqref{int_contri}) is
$\sim \{10^{-4}-10^{-3}\}\times\Delta g\langle\sigma^z\rangle$ and
will dominate when $\Delta g > 10^{-3}$. We return to this later.
The mobility edge channel gives rise to very concrete formulae for
the behavior of the magnetoresistance and Hall effect. So now we
should examine the hopping magnetoresistance and then we may be able
to decide which mechanism dominates in each range of B-field and
temperature.
\section{Magnetoresistance}
\subsection{Hopping Magnetoresistance}
The relative importance of the hopping and band edge weak
localization contributions could, in principle, be tested by
measuring the corresponding magnetoresistance. We shall now show
that these are very different, both in structure and in order of
magnitude. It is surprising that magnetoresistance was not measured
by Allen \emph{et al}.\cite{Allen}. The magnetoresitance in the
apparently \emph{hopping like regime} was measured in a similarly
low doped (2\%) system by Van Esch \emph{et al}.\cite{Van_Esch1},
who found a strong negative magnetoresistance, which they
interpreted as being due to the orbital expansion of the localized
states in a B-field. This, the authors suggested, is caused by the
competition between the spin Zeeman energy and the antiferromagnetic
hole-Mn coupling. The external field is, however, far too small to
interfere with the strong antiferromagnetic $J_{pd}$ coupling of
$\sim 0.5$eV\cite{Jungwirth1}. The Van Esch \emph{et
al}.\cite{Van_Esch1} magnetoresistance data actually look very
similar to those obtained for metallic films by Matsukura \emph{et al}.\cite{Matsukura}.\\
\\
Let us now examine the complete hopping magnetoresistance and then
evaluate whether it can explain the available data\cite{Van_Esch1}
in the insulating regime. The theory will also allow us to make
predictions concerning what should be the behavior of the hopping
magnetoresistance.\\
\\
Specifically, hopping spin and orbital magnetoresistance (MR) do
exist\cite{Mell_Stuke,Kochman,Bobbert}, and have been discussed by
Movaghar and Schweitzer\cite{Movaghar_Schweitzer} as applied to
a-Si. Hopping magnetoresistance usually has a positive contribution
from orbital shrinking of wavefunctions at very  high field, $B>1$T
(Mikoshiba effect). This is a one-body effect. Localized states also
have a finite Hubbard $U$, so there is also a two-body contribution
to the spin hopping magnetoresistance. An electron from below the
Fermi level can, in general, hop to an empty state, or one which is
already occupied by another electron below. The doubly occupied
level is shifted up by the Hubbard $U$ and, if $U$ is not too large,
this takes some of the doubly occupied states to the Fermi level.
The jump to the occupied state takes place in the triplet or singlet
configuration. The triplet hop being forbidden, the three triplet
states would have to first convert into singlets. This process
depends sensitively on (small) transverse ($B_x$,$B_y$) magnetic
fields within the system\cite{Movaghar_Schweitzer}.  The
triplet-to-singlet transformation can also occur during the jump via
the spin-orbit coupling $\lambda\textbf{l}\cdot\boldsymbol{\sigma}$.
We can write the totality of effects as\cite{Movaghar_Schweitzer}
\begin{equation}\label{hop_magn_Mo_Sch}
\begin{split}
     &\sigma_{xx}^{hop}(B_z,T) =
    \sigma_0\text{e}^{  -\left(\d{T_0}{T}\right)^{\d{1}{4}}\left( 1 + \xi\left(\d{eB_z}{\hbar}\right)^2 \right)
    }\\
    &\times\left[ N_e + N_S + N_T\left( \d{\nu_{so}}{\nu_0} + \d{ \left(\mu_B\d{\langle B_{int}(0) \rangle}{\hbar}\right)^2\tau^2 }{(\omega_c\tau)^2 + 1} \right) \right].
\end{split}
\end{equation}
In Eq.~\eqref{hop_magn_Mo_Sch}, $\xi$ is a quantity that will be
given below. $N_e$, $N_S$ and $N_T$ are probabilities that the hop is
to a previously empty, singlet or triplet state. $N_S$ and $N_T$
depend upon the magnetic field as $N_S =
\d{1-\langle\sigma_z\rangle}{4}$ and $N_T =
\d{3+\langle\sigma_z\rangle}{4}$. The direct spin-orbit assisted
triplet-to-singlet hop is proportional to $\d{\nu_{so}}{\nu_0}$
where $\nu_0$ is the hopping (singlet) prefactor. The internal
random fluctuating magnetic fields $B_{int}(t)$ which allow spin
admixtures to occur are mainly due to fluctuations in $g^{xz}$,
$g^{yz}$ and the anisotropic part of the dipole-dipole interaction.
For weak coupling, we have
\begin{equation}\label{av_Hint_squared}
    \langle B_{int}^2 \rangle = \langle B_{dip}^2 \rangle +
    \d{1}{2}\left[ \left|\Delta g_{xz}B_z^{ext}\right|^2 + \left|\Delta g_{yz}B_z^{ext}\right|^2 \right].
\end{equation}
The dipolar internal fields $\langle B_{dip}^2 \rangle^{1/2}$ are of
the order of the ESR linewidths smaller than $10$mT, the
$g$-admixture is $\sim 10^{-3}$. Saturation occurs when the
frequencies corresponding to Eq.~\eqref{av_Hint_squared} are
comparable with typical hopping frequencies. A similar model has
been used recently to explain the
magneto-electro-luminescence in polymers\cite{Sheng}.\\
\\
The Mikoshiba effect, mentioned earlier, is spin independent and
changes the phonon assisted hopping rate as
\begin{equation}\label{hop_rate}
    W_{ij} = W_{ij}(0)\text{e}^{-2\alpha R_{ij} - \d{R_{ij}^3}{6\alpha}\left(\d{eB}{\hbar}\right)^2 - \d{\varepsilon_j-\varepsilon_i}{k_BT}
    },
\end{equation}
averaged over the angle between the B-field and \textbf{R} and with
$\varepsilon_j > \varepsilon_i$. At optimum hopping length,
$R_{ij}^3$ goes to $\langle (\alpha R)^3\rangle =
\left(\d{3}{8}\right)^{\d{3}{4}}\left(\d{T_0}{T}\right)^{\d{3}{4}}$
and $\d{|\varepsilon_j-\varepsilon_i|}{k_BT}$ goes to
$\d{1}{4}\left(\d{T_0}{T}\right)^{\d{1}{4}}$. Inserting these terms
in Eq.~\eqref{hop_rate} gives us $\xi = \left( \d{T_0}{T}
\right)^{\d{1}{2}}\left(\d{3}{8}\right)^{\d{3}{4}}\d{1}{6\alpha^4}$
in Eq.~\eqref{hop_magn_Mo_Sch}. To lowest order, we may neglect the
change in $T_0$ caused by $B_z$. There is also an increase of the
energy of the localized state. This shift lowers the ionization
energy to the mobility edge which, in first order perturbation
theory is
\begin{equation}\label{mob_edge}
    \varepsilon_i(B_z) = \varepsilon_i + \d{ (eB_z)^2 }{2\alpha_i^2m},
\end{equation}
where $\alpha_i$ is the inverse localization length of the $i^{th}$
level. Finally, there is also a small quantum interference
magnetoresistance contribution due to the fact that the hopping
carrier could also virtually visit the second best neighbor site on
the way to its best neighbor. This contribution has been analyzed in
detail by Schirmacher\cite{Schirmacher}. The spatial part of the
transfer rate is, to first order, proportional to
\begin{equation}\label{trans_rate}
\begin{split}
    &\left| T_{ij}(B)\right|^2 =\\ &\left|t_{ij}^0\text{e}^{ -\d{ie}{\hbar}\textbf{B}\cdot (\textbf{R}_i\times\textbf{R}_j)
    } + \sum_l\d{t_{il}^0t_{lj}^0}{\varepsilon_i -
    \varepsilon_l}\text{e}^{ -\d{ie}{\hbar}\textbf{B}\cdot (\textbf{R}_i\times\textbf{R}_l +
    \textbf{R}_l\times\textbf{R}_j)}\right|^2.
\end{split}
\end{equation}
This now includes the direct transfer, and the interference term
from a third site. This means that the
Miller-Abrahams\cite{Miller_Abrahams} hopping spatial factor,
$|t_{ij}^0|^2$, should be replaced by Eq.~\eqref{trans_rate} which
is a small but magnetic field dependent factor. One need take only
one term in the $l$ sum, the second nearest neighbor term. The
optimized second neighbor $2\alpha R_{il}$ and $\d{\varepsilon_l -
\varepsilon_i}{k_BT}$ will also scale as $ \gamma\left( \d{T_0}{T}
\right)^{\d{1}{4}} $ but with a constant factor $\gamma$, which
instead of the first neighbor values, 3/4 and 1/4 respectively, will
be somewhat larger, $\sim 39/32$ for the spatial length and 13/32
for the energy. We have already used this for the hopping Hall
effect (Eq.~\eqref{mob_xy}). The sign depends upon the geometry of
the triad and can be positive or negative.\\
\\
We conclude that in Eq.~\eqref{hop_magn_Mo_Sch}, we have to replace
$\sigma_0$ by $\sigma_0\left\langle
\left|\d{T_{ij}(B_z)}{t_{ij}}\right|^2 \right\rangle$ where
$\langle\rangle$ is the spatial/energy average over the best second
neighbor (see Refs.~\onlinecite{Movaghar2}, \onlinecite{Gruenewald} and Refs therein for the definition of best second neighbor). Keeping the best third neighbor
$l$ and assuming $R_i=0$, we obtain a quantum interference
contribution
\begin{equation}\label{quan_contri}
    \Delta\sigma_0(B) = \sigma_0\left\langle
    \d{2t_{il}t_{lj}}{t_{ij}(\varepsilon_i-\varepsilon_l)}\left[\cos\left(
    \d{e}{\hbar}\textbf{B}\cdot (\textbf{R}_l\times\textbf{R}_j)\right)-1\right]
    \right\rangle.
\end{equation}
The average in Eq.~\eqref{quan_contri} is over the percolation network and gives a negative magnetoresistance,
according to Schirmacher\cite{Schirmacher}. In general, one has to note that all hopping contributions below
$1$T are at best a few per cent, and nowhere near the large values
observed by Van Esch \emph{et al}.\cite{Van_Esch1}. This completes
the hopping analysis. Now consider the contribution to the
magnetoresistance coming from excitations to the band edge.
\subsection{The \emph{orbital} magnetoresistance from the region near the mobility edge}
Following Khmelitskii \emph{et al}.\cite{Khmelnitszkii} and Movaghar
and Roth\cite{Movaghar_Roth}, we have for holes and small B-fields,
the following shift for the mobility edge:
\begin{equation}\label{EFminusEC}
    \left|\varepsilon_f(B_z)-\varepsilon_c(B_z)\right| \sim \varepsilon_f(0) - \varepsilon_c(0)\left[ 1 + \kappa\left(\d{\hbar\omega_c}{\varepsilon_c(0)}\right)^{\d{1}{2}}\right],
\end{equation}
where $\kappa$ is a constant of order 1 and $\omega_c$
is the cyclotron frequency of the holes. In Eq.~\eqref{EFminusEC}, we have neglected the magnetic field shift in the Fermi energy compared to that of the mobility edge. This result is valid for
fields at which $\hbar\omega_c < \hbar /\tau$, cyclotron energy is
less than the scattering broadening, and gives, from
Eq.~\eqref{EFminusEC}, an
$\text{exp}\left[\d{\kappa\left[\hbar\omega_c\varepsilon_c(0)\right]^{\d{1}{2}}}{k_BT}
\right]$ increase in the conductivity over a limited field range.\\
\\
The above describes the well known orbital delocalization effect.
The fully self-consistent shift in the effective activation energy,
i.e. the exact value of the one-body
$\varepsilon_f(B_z)-\varepsilon_c(B_z)$ shift, which includes the
ferromagnetic effects, and high B-fields is not known and this would
need a detailed study of localization in a disordered ferromagnet.
However, it can be inferred that in the present case, field induced
magnetic alignment is going to lower disorder and thus increase the
range of delocalized states by shifting $\varepsilon_c(B_z,M_z)$ up
with $M_z$ and $\varepsilon_f(B_z)$ down with $B_z$. At low
temperatures, the magnetism can be saturated at relatively low
fields and will not change the electronic structure very much below
$T_c$.
\section{Summary of contributions, including spin disorder scattering}
\begin{figure}
        \begin{center}
            \includegraphics[scale=0.4]{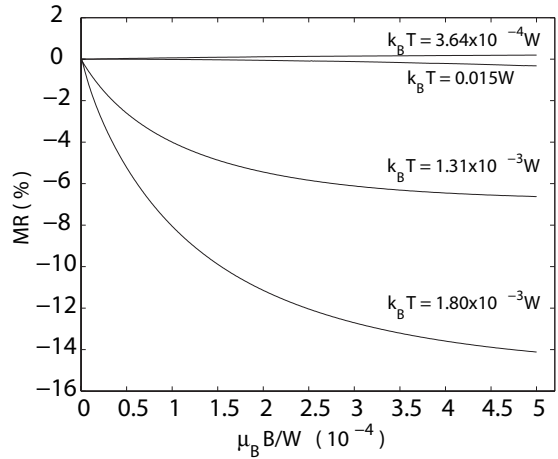}
        \end{center}
        \caption{Relative magnetoresistivity in the CPA for carrier concentration $\text{p} = 0.1x$
       for impurity concentration $x = 0.053$ and $J_{pd} = 0.35W$, with $T$ as a parameter. $W$ correspond to half the bandwidth. Figure taken from Ref.~\onlinecite{Arsenault1}. This is the expected behavior just below (hole) the mobility edge}
            \label{fig:MR_B_J035x053EM0}
\end{figure}

\begin{figure}
        \begin{center}
            \includegraphics[scale=0.4]{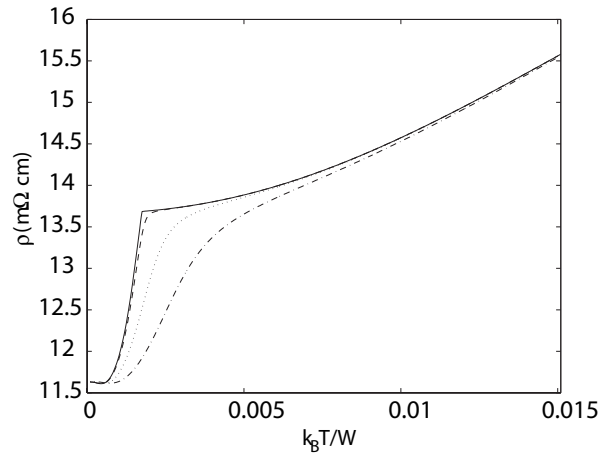}
        \end{center}
        \caption{Resistivity as a function of temperature for various
magnetic fields. Others parameters are $x$ = 0.053, $E_{NM} = E_M =
0$, $J_{pd} = 0.35W$ and $\text{p} = 0.1x$. $W$ correspond to half
the bandwidth. The solid line is the zero field result, the dashed
line is for $\mu_BB = 1\text{x}10^{-5}W$, the dotted line is for
$\mu_BB = 1\text{x}10^{-4}W$ and the dashed-dotted line is for
$\mu_BB = 3\text{x}10^{-4}W$. Figure taken from
Ref.~\onlinecite{Arsenault2}}
            \label{fig:resis_B0x053EM0_B_T}
\end{figure}

We now have a theory which allows us to calculate, in principle, the
resistance, magnetoresistance, and Hall coefficient when the Fermi
level is in the region of localized states. We have also included
the band region at, or just above, the mobility edge including the
spin scattering effects which can be described by the CPA (Coherent
Potential Approximation). This is discussed in
Appendix~\ref{appen_3}. The Hall conductivity has already been given
by Eq.~\eqref{3contri_trans_conduc}.\\
\\
Consider first the complete longitudinal conductivity in the
presence of a magnetic field $B_z$ and including orbital and spin
contributions from localized and delocalized states, this is given
by
\begin{equation}\label{3contri_conduc}
\begin{split}
    \sigma_{xx} &= \d{
    e^2}{\pi}\int_{\varepsilon_b}^{\varepsilon_c(B_z)}d\varepsilon\left(-\d{\pd
    f(\varepsilon)}{\pd\varepsilon}\right)\rho_{edge}(\varepsilon)D_0\left|\d{\varepsilon-\varepsilon_c(B_z)}{\varepsilon_c(B_z)}\right|\\
    &+ \int_{-\infty}^{\varepsilon_b}d\varepsilon\left( -\d{\pd
    f(\varepsilon)}{\pd\varepsilon}
    \right)\rho_{band}(\varepsilon)D_{CPA}(\varepsilon,B_z)\\ &+
    \sigma_{xx}^{hop}(B_z),
\end{split}
\end{equation}
where $\varepsilon_b$ is the energy above which the CPA stops being
applicable. This energy is close to, but a little bit below, the
mobility edge (we are considering hole). This expression includes
the band edge (first term), the high mobility band term (second
term), and the third term, which is the hopping contribution. In the
first, or band edge term, the magnetic field dependence is basically
due to the B-field induced delocalization, and not to spin. In the
second, or band term, which takes over at higher temperatures, the
magnetoresistance is mainly due to the spin-disorder scattering, and
is negative, because the magnetic field reduces the disorder, and
thus reduces the scattering rate. Figs.~\ref{fig:MR_B_J035x053EM0}
and ~\ref{fig:resis_B0x053EM0_B_T} are examples of this behavior.
Unfortunately, we do not have the analytical form for this term, but
we have the numerical CPA results shown in
Fig.~\ref{fig:MR_B_J035x053EM0} and
Fig.~\ref{fig:resis_B0x053EM0_B_T}.
Also, from Eq.~(15) of Ref.~\onlinecite{Van_Esch1}, we can create an
approximation to the CPA from perturbation theory. By defining, the
$pd$ scattering time as
\begin{eqnarray}\label{pd_scatt}
&&\d{1}{\tau_{pd}}
=\d{J_{pd}^2\rho_{band}(\varepsilon)}{\hbar}\Bigg[ S(S+1) -
S^2\left(\d{M_z}{M_s}\right)^2\nonumber\\
&-&S\left(\d{M_z}{M_s}\right)\tanh\left(\d{3T_cM_z}{2TS(S+1)M_s}\right)
\Bigg],\nonumber\\
\end{eqnarray}
the CPA diffusivity can be written
\begin{equation}\label{CPA_Diff}
    D_{CPA} \sim
    v(\varepsilon)^2\d{\tau_{pd}\tau_{e}}{\tau_{pd}+\tau_{e}},
\end{equation}
where $\tau_{e}$ is the non magnetic relaxation time and
$v(\varepsilon) = \sqrt{\d{2\varepsilon}{m^*}}$ is the band velocity
at energy $\varepsilon$. Eqs.~\eqref{pd_scatt} and ~\eqref{CPA_Diff}
could be interpreted as a first semi-analytical approximation to the
CPA result (see Fig.~\ref{fig:MR_B_J035x053EM0} for a numerical evaluation). The spin-hopping term is given by \eqref{hop_magn_Mo_Sch}.\\
\\
The density of states has been divided into three sections
corresponding to the region near the Fermi level (hopping), the
region at the band edge (Eq.~\eqref{dos_exp}), and the delocalized
region just below $\varepsilon_c$, in the CPA regime.\\
\\
We have established that when the Fermi level is in the region of
localized states, both band edge and hopping will, in general,
contribute to transport. The question is to what extend is it one or
the other? This obviously depends upon temperature. But it has been
argued that if the observed conductivity scales as Mott's law, then
the explanation cannot be anything else but VRH (variable range
hopping), for all the transport coefficients. There are some facts
that speak against this simple interpretation. The thermopower data
are not hopping-like\cite{Osinniy}. Our estimates show that the
AHE\cite{Allen}, can, at very best, only be due to hopping at very
low temperatures. But the Hall data, thermopower and
magnetoresistance are all on different samples, so one has to be
very careful and rigorous, and examine all potential contributions.\\
\\
Indeed, with the usual assumption of exponential density of states
(Eq.~\eqref{dos_exp}), the first term in Eq.~\eqref{3contri_conduc}
will give an activated band edge transport behavior, so, in this
limit, we have the complete result,
\begin{widetext}
\begin{equation}\label{sig_bandedge_exp}
\begin{split}
\sigma_{xx} &=
e^2D_0\rho_0(\varepsilon_f(B_z))\d{k_BT}{\varepsilon_c(B_z)}\text{e}^{-\d{\varepsilon_f(B_z)
- \varepsilon_c(B_z)}{k_BT}+
\d{\varepsilon_c(B_z)}{\varepsilon_0}}+\sigma_{xx}^{hop}(T,B_z)\\
&\equiv \sigma_{xx}^{BE}(T,B_z)+\sigma_{xx}^{hop}(T,B_z).
\end{split}
\end{equation}
\end{widetext}
Using Eq.~\eqref{int_contri} and replacing the quantum diffusivity $\hbar/m^*$ with the normal diffusivity, we have
\begin{widetext}
\begin{equation}\label{sigxy_bandedge_exp}
\sigma_{xy} = \sigma_{xx}^{BE}(T,B_z)\left[
\d{eB_z\tau(\varepsilon_c(B_z))}{m^*} + \Delta
g(\varepsilon_c(B_z))\langle\sigma_z\rangle_{\varepsilon_c} \right]
+\sigma_{xy}^{hop}(T,B_z).
\end{equation}
\end{widetext}
If we had kept the original form of Eq.~\eqref{int_contri}, the dependence of the Hall coefficient ($R_H$) on the resistance would
be resistance squared as observed in most ferromagnets.\\
\\
To complete the analysis we also need Eq.~\eqref{EFminusEC}. The
conclusion one can draw by looking at
Eqs.~\eqref{sig_bandedge_exp},~\eqref{sigxy_bandedge_exp} and
\eqref{EFminusEC} is that, as temperature increases, an activated
law should be observed, with a large $\text{exp}\left(
\kappa\varepsilon_c\sqrt{\hbar\omega_c/\varepsilon_c}/k_BT \right)$
increase in conductance with B (negative magnetoresistance).\\
\\
A word of caution: the shift in mobility edge with B-field as
applied to electrons or holes assumes that the excited state
wavefunction has time to build an Anderson level with all its
features\cite{Khmelnitszkii}. This is not necessarily true for
highly excited states. As an example, take a-Si:H where no such
negative MR is observed despite activation above the mobility edge.
In the present case, the mobility edge is very close to the Fermi
level so that it is reasonable to assume that this mechanism is also
operating.\\
\\
Van Esch \emph{et al}.\cite{Van_Esch1} fit their data to a $\left(
\d{\hbar\omega_c}{k_BT} \right)^{1/3}$ law. But most often a Mott
law is observed in conduction. Osinniy \emph{et al}.\cite{Osinniy}
have analyzed the thermopower in low doped insulating samples and
observed an activated law (see Appendix~\ref{appen_thermo} for
thermopower). The resistance they measured also looks activated,
although the authors do not explicitly state it. So their data would
agree with the band edge formulae. From this, one would have to
conclude that both band edge transport and hopping is observed,
depending upon the sample and the transport coefficient. But when we
examine the order of magnitude of the hopping magnetoresistance we
have, at most, a few percent, even at very high fields. The data of
Van Esch \emph{et al}.\cite{Van_Esch1} on magnetoresistance give us
a strong clue as to what might be happening. They have a few hundred
percent $MR$ but, with Mott like conductivity. So, finally, we have
one last question to examine. We need to investigate what other
mechanism can give us a Mott-like law for conduction but a giant
magnetic band-like magnetoresistance and AHE. Let us examine this
possibility before we come to a final decision on the mechanism.
\subsection{Mott law from correlation holes or granularity}
The influence of correlations was treated by Anglada \emph{et
al}.\cite{Anglada}, and more recently, by Dunford \emph{et
al}.\cite{Dunford}, in the context of transport in polymers,
granular systems and nanoparticle composites. The point is that in
highly correlated systems, the density of excited states at the band
edge need not, and in general will not, be exponential. Coulomb
energies and spin fluctuations can modify the excited state
spectrum. Though the exponential structure for $\rho(\varepsilon)$
is reasonably well established in systems which are weakly
correlated, such as nonmagnetic amorphous
semiconductors\cite{Dyalsingh_Kakalios}, one knows that in electron
and spin glasses, this is not the case, and final state interactions
are important. When an empty state just above the Fermi level is
filled by an electron excited from the valence band, the delocalized
hole created in the valence band is not immediately neutralized. The
hole which is created recombines elsewhere in the material, with an
electron from below the Fermi level, and in this way, charge has
been transported. The process of the electron separating itself from
its hole is complex in a highly correlated spin system such as
GaMnAs. Here, it is the carriers which cause the spin alignment in
the first place. We can expect the density of states for excitations
in such a system to be more sharply cut off at small energies. We
shall see that the proposed structure will reproduce a coulomb glass
type behavior with temperature\cite{Moebius}. A similar situation is
encountered in granular systems\cite{Movaghar_Roth,Anglada} where
Mott-like laws are observed as a result of inter-grain thermally or
fluctuation-assisted tunneling\cite{Sheng1}. Granularity is also
expected in these materials. Indeed we can simulate this behavior
with a density of states per volume of the form
\begin{equation}\label{dos_edge_gran}
    \rho_{edge}(\varepsilon) =
    \rho_{band}\text{e}^{-\left(\d{\varepsilon_g}{|\varepsilon_f-\varepsilon|+\varepsilon_l}\right)^{\nu}},
\end{equation}
characteristic  of correlated or granular systems, with $0 < \nu <
1$ and where $\varepsilon_g$, $\varepsilon_l$, are constants which
measure steepness and the density of states at the Fermi level. Finally, $\rho_{band}$ is a constant.\\
\\
Knowing the density of states at the band edge, we can now proceed
to evaluate the Hall effect, longitudinal conductivity and
magnetoresistance at the band edge. To see how we can get a Mott
type law from extended states, we use the band edge term of
Eq.~\eqref{3contri_conduc} and Eq.~\eqref{dos_edge_gran} to get
\begin{widetext}
\begin{equation}\label{sig_granu}
    \sigma_{xx} =
\d{e^2}{k_BT}\int_{-\infty}^{\varepsilon_c(B_z)}d\varepsilon\rho_{band}D_0\left|\d{\varepsilon_c(B_z)-\varepsilon}{\varepsilon_c(0)}\right|\text{e}^{-\d{\left|\varepsilon_f(B_z)-\varepsilon\right|}{k_BT}
-
\left(\d{\varepsilon_g}{|\varepsilon_f(B_z)-\varepsilon|+\varepsilon_l}\right)^{\nu}}.
\end{equation}
\end{widetext}
A steepest descent evaluation of Eq.~\eqref{sig_granu} at low $T$
gives for $\nu=1/2$, an
$\text{exp}\left[-\left(\d{T_0}{T}\right)^{1/3}\right]$ law (see
Appendix~\ref{appen_steep}). Here, the Mott-like laws are a result
of the pseudo-gap in the density of states or a depletion of low
energy excitations created by long range correlations.
\subsection{Quantitative comparison: VRH versus delocalized band edge conduction}
To asses the relative importance of the various channels, let us
compare some values starting with $\sigma_{xx}$. The band edge
conductivity with a correlation hole density of states evaluated
using steepest descent (Appendix~\ref{appen_steep}), and the usual
hopping conductivity, would give us (with $\nu = 1/3$)
\begin{widetext}
\begin{equation}\label{sig_band_steep}
    \sigma_{xx}(band) =
    \d{e^2}{\varepsilon_c(B_z)k_BT}D_0\rho_{band}\Big[\varepsilon_{max}-[\varepsilon_f(B_z)-\varepsilon_c(B_z)]\Big]^2\text{e}^{-\d{3}{2}\left(\d{\varepsilon_g}{k_BT}\right)^{1/4}},
\end{equation}
\end{widetext}
where
\begin{equation}
    \varepsilon_{max} =
    \left(\nu k_BT\varepsilon_g^{\nu}\right)^{\d{1}{1+\nu}}.
\end{equation}
With Eq.~\eqref{EFminusEC}, for holes, we obtain\\
\begin{widetext}
\begin{equation}
    \sigma_{xx}(band) =
    \d{e^2}{\varepsilon_c(0)k_BT}D_0\rho_{band}\left(\varepsilon_{max}-\left[\varepsilon_f(0)-\varepsilon_c(0)\left( 1 + \kappa\left(\d{\hbar\omega_c}{\varepsilon_c(0)}\right)^{\d{1}{2}} \right)\right]\right)^2\text{e}^{-\d{3}{2}\left(\d{\varepsilon_g}{k_BT}\right)^{1/4}},
\end{equation}
\end{widetext}
and
\begin{equation}\label{mag_resis}
    \d{\Delta\sigma_{xx}}{\sigma_{xx}} = \d{2\varepsilon_{max}\varepsilon_c(0)\kappa\left( \d{\hbar\omega_c}{\varepsilon_c(0)} \right)^{\d{1}{2}}}{\Big(\varepsilon_{max}- [\varepsilon_f -
    \varepsilon_c(0)]\Big)^2}.
\end{equation}
For the Hall conductivity, we obtain, to a good approximation
$(\nu=1/3)$, for holes 
\begin{equation}\label{sigxy_band_steep}
\begin{split}
    &\sigma_{xy}(band) =\\ &\sigma_{xx}(band)\left[\d{eB_z\tau(\varepsilon_{max})}{m^*}  + \langle \Delta g(\varepsilon_{max})
    \rangle\langle\sigma_z\rangle_{\varepsilon_{max}}
    \right].
\end{split}
\end{equation}
Whereas for VRH hopping (without the magnetoresistance terms) we
have
\begin{equation}
\sigma_{xx}(hopping) = \d{e^2}{4\alpha^2}\rho(\varepsilon_f)\left[
\d{3}{4}\left(\d{T_0}{T}\right)^{1/4}
\right]^{2}\nu_0\text{e}^{-\left(\d{T_0}{T}\right)^{1/4}}.
\end{equation}
It is understood that the $T^{1/4}$ factors are chosen to be the
same so that this determines the density of states parameters
$\varepsilon_g$. Thus, $(3/2)^4\varepsilon_g/k_B = T_0$. The
quantities, $\rho_{band} \sim 10^{20}-10^{21}$/cm$^3$eV, $D_0
\sim 10^{-4}$m$^2$/s, $\rho(\varepsilon_f) \sim
10^{17}-10^{19}$/cm$^3$eV. $T_0$ or $\varepsilon_g$ is fixed by
experiment. From the data of Allen \emph{et al}.\cite{Allen}, we
extract roughly $T_0=2.4\times10^5$, $T_0 = 24\times
2.7\alpha^3/k_B\rho(\varepsilon_f)$. The localization radius
corresponds to a binding energy of 110 meV which is $\sim 3nm$ for a
light hole. The corresponding $\rho(\varepsilon_f)\sim 3\times
10^{19}$/cm$^3$eV extracted is a very large number and normally
above the value required for delocalization.\\
\\
One can see that the band pseudo-hopping expression can easily
account for the dc conduction. We see that by selecting $\nu$, one can create an
$\text{e}^{-\left(\d{T_0}{T}\right)^{1/2}}$ or $\text{e}^{-\left(\d{T_0}{T}\right)^{1/4}}$
behavior without variable range hopping. Thus, if some experiments give one or the other law,
it only means that the effective density of states is different. This has nothing to do with VRH. Furthermore, from
Eqs.~\eqref{EFminusEC} and \eqref{mag_resis}, it gives the
correct order of magnitude, the right sign and nearly the exact form of
the magnetoresistance, whereas the hopping magnetoresistance is
\emph{nowhere near the right order of magnitude and structure}. The
magnetism and large magnetoresistance observed by Van Esch \emph{et
al}.\cite{Van_Esch1} at very low temperature suggest that
inhomogeneity and clustering play an important role. Hopping
magnetoresistance would, however, seem to be the correct explanation for the magneto-electroluminescence data in the work of Mermer
\emph{et al}.\cite{Mermer} where effects are a few percent for
B-fields of mT, as in the magnetoresistance and photoconduction of
a:Si\cite{Mell1,Mell_Stuke,Movaghar_Schweitzer}.\\
\\
Finally, the same optimization applied to the Hall effect gives,
from Eq.~\eqref{sigxy_band_steep}
\begin{equation}\label{RH_band_steep}
    R_H = \left[\d{e\tau(\varepsilon_{max})}{m^*} + \d{\left\langle\Delta g(\varepsilon_{max})\right\rangle
    \langle\sigma_z\rangle_{\varepsilon_{max}}}{B_z}\right]\d{1}{\sigma_{xx}(band)}.
\end{equation}
In this form, we can see that the anomalous term easily competes
with the normal term since $\tau\sim 10^{-14}$s, $\langle\Delta
g\rangle \sim 10^{-2}-10^{-1}$ and
$\langle\sigma_z\rangle_{\varepsilon_{max}} \sim 0.1-1$. Such a
large effect is very difficult to achieve with hopping interference
paths over third neighbors, even though the square root dependence
on resistance predicted by hopping theory is actually closer to
experiment than the linear one predicted here in
Eq.~\eqref{RH_band_steep}. The thermopower and resistance data by Osinniy \emph{et al}.\cite{Osinniy} also suggest band edge conduction with the usual
exponential density of state (Eq.~\eqref{dos_exp})(see
Appendix~\ref{appen_thermo}).
\section{Discussion and Conclusion}
Our analysis suggests that hopping dominates only the very low
temperature region, and even then, it is most likely
cluster-to-cluster hopping so that energy differences will depend on
the cluster magnetism and not just on disorder. For temperatures above
$T\sim 10K$, we believe that the Hall coefficient, magnetoresistance
and thermopower are due to band edge conduction. The observed
\emph{Mott-like} laws are most likely caused by Coulomb
correlations. In band edge transport, any value of $\nu$ between 0 and 1 can be expected.
The stronger the correlations, the smaller is $\nu$. Experiment does not give a unique
value of $\nu$, but it is always band edge conduction which is observed. In band conduction, the carriers are subjected to
incomparably stronger magnetoresistance and Hall forces then in hopping conduction. This gives
rise, in particular, to the anomalous Hall effect. In Appendix~\ref{appen_thermo}, we show that the
conductivity and thermopower data of Osinniy \emph{et
al}.\cite{Osinniy} also agree with the model of delocalized band
transport near the mobility edge. It appears that the temperature
dependence of the transport data for insulating samples depends, to
some extent, on sample preparation. When the totality of
available data in the insulating regime of GaMnAs, except those at
the lowest of temperatures, are examined, it appears to be either activated
transport or correlation band transport which is the dominant
mechanism. There are strong indications that Mott hopping plays a
secondary role at this temperature and ferromagnetic concentration
range. There is, rather, a transfer process by way of which a
localized hole is annihilated and created elsewhere via thermal
fluctuations and is a band carrier in the intermediate state where
it lives long enough to be subject to the band Hall forces and spin
disorder scattering. In other words, the long range band transfer
mechanism that gives rise to the ferromagnetism in the material is
also the mechanism that determines the transport. The charge
transfer process is predominantly over the band with a
Shklovskii-Efros\cite{Shlovskii_Efros} type correlation band edge.
The picture we propose is similar to that of Durst \emph{et
al}.\cite{Durst} for this regime. It is the dynamics of the shared
spins, when looked at as real or virtual band excitations, which
determines the magneto-and Hall transport. At very low temperatures
there must be a region in which hopping at the Fermi level dominates. It would be interesting to look for it experimentally. We predict
that in this range the MR and Hall effect are given by
Eq.~\eqref{hop_magn_Mo_Sch} together with Eq.~\eqref{quan_contri}
and Eq.~\eqref{rho_xy_div_B} and are much smaller. Very recently,
weak localization in GaMnAs has been
observed\cite{Rokhinson,Neumaier}. This regime is not exactly the
one we study here ($\varepsilon_f$ is a somewhat farther from the
mobility edge in the insulating regime than in the weak localization
regime) but their results are in agreement with our model.\\
\\
We have presented a reasonably complete theory of transport in
magnetic semiconductors of the GaMnAs type, when the Fermi level is
in the region of localized states, in the presence of disorder and
magnetism. The data available is, unfortunately, sparse and
scattered and there does not seem to be a complete set of data on
any one set of samples. The formulae presented here have a wide
range of applications. The principal result of the paper is that we conclude that transport
in Mn doped GaAs is due to activated transport rather than variable range hopping. Finally, the possibility of magnetic-cluster-to-cluster hopping
and tunneling should be taken seriously and investigated in detail.
The resulting magnetoresistivity and Hall effect properties of such
systems are not easy to guess.
\begin{acknowledgments}
The authors gratefully acknowledge financial support from the
Natural Sciences and Engineering Research Council of Canada (NSERC)
during this research. P.D. also acknowledges support from the Canada
Research Chair Program. L.-F. A. gratefully acknowledge financial support from
Pr. André-Marie Tremblay during the writing of the manuscript.
\end{acknowledgments}
\appendix
\section{Steepest Descent Integral}\label{appen_steep}
Using a steepest descent approach with Eq.~\eqref{sig_granu} to
calculate the conductivity and Hall effect, and neglecting the
energy dependence, apart from the density of states and Fermi
function, we have an integral of the type (assume $\varepsilon_l=0$)
\begin{equation}\label{I1}
    I \sim\int_{\varepsilon-\varepsilon_f}^\infty
    d\varepsilon\text{e}^{-\d{\varepsilon}{k_BT} -
    \left(\d{\varepsilon_g}{\varepsilon}\right)^{\nu}}G(\varepsilon).
\end{equation}
We have changed the sign of the integral from "holes to electrons"
for convenience. The maximum of the integrand, neglecting the energy
dependence of G is at
\begin{equation}\label{Emax}
    \varepsilon_{max} \sim \left[\nu
    k_BT\varepsilon_g^{\nu}\right]^{\d{1}{1+\nu}},
\end{equation}
provided that $\varepsilon_{max} > |\varepsilon_c-\varepsilon_f|$.
Clearly this condition and $\varepsilon_l=0$ are not always
satisfied, and depend on the material and temperature. When this
limit is not valid, the integral is controlled by the lower limit
and the conductivity is activated i.e.
\begin{widetext}
\begin{equation}\label{I2}
    I\sim\int_{\varepsilon-\varepsilon_f}^\infty d\varepsilon\text{e}^{-\d{\varepsilon}{k_BT} -
    \left(\d{\varepsilon_g}{\varepsilon+\varepsilon_l}\right)^{\nu}}G(\varepsilon)
    \sim k_BTG(\varepsilon_f-\varepsilon_c)\text{e}^{ -
    \left(\d{\varepsilon_g}{(\varepsilon_f-\varepsilon_c)+\varepsilon_l}
    \right)}.
\end{equation}
\end{widetext}
When it is valid, the best value can be substituted back, and gives
a temperature dependence of the form
\begin{widetext}
\begin{equation}\label{I3}
    I\sim \d{\varepsilon_{max}G(\varepsilon_{max})}{2}\text{e}^{-\d{\varepsilon_{max}}{k_BT} -
    \left(\d{\varepsilon_{g}}{\varepsilon_{max}}\right)}\sim
    \text{e}^{-\left[\left(\d{\varepsilon_{g}}{k_BT}\right)^{\d{\nu}{1+\nu}}(\nu^{\d{\nu}{1+\nu}} +
    \nu^{\d{-\nu}{1+\nu}})\right]},
\end{equation}
\end{widetext}
which is a Mott-like law, but without VRH.
\section{Thermopower}\label{appen_thermo}
$E_{xx}$, the energy transported and $S$, the thermopower without
interaction effects, are given by Movaghar and
Schirmacher\cite{Movaghar_Schirmacher}:
\begin{eqnarray}\label{Exx}
    E_{xx} = \int_{-\infty}^{\varepsilon_c}&&d\varepsilon\left(-\d{\pd
    f(\varepsilon)}{\pd\varepsilon}\right)\rho_{edge}(\varepsilon)D_0\left| \d{\varepsilon - \varepsilon_c}{\varepsilon_c}
    \right|\left(\varepsilon-\varepsilon_f\right)\nonumber\\ &+&E_{xx,hop}
\end{eqnarray}
and
\begin{equation}\label{S}
    S = \d{eE_{xx}}{\sigma_{xx}T}.
\end{equation}
In the correlation model with $\nu=1/3$ and for low enough
temperatures, given by $\varepsilon_{max} >
\left|\varepsilon_c-\varepsilon_f\right|$
\begin{equation}\label{S1}
    S = \d{\left(\nu
    k_BT\varepsilon_g^{\nu}\right)^{\d{1}{1+\nu}}}{eT}.
\end{equation}
This yields a $T^{-1/4}$ law and a magnetic field dependence, which
is part of the Coulomb glass energy $\varepsilon_g(B_z)$ (effective
Mott $T_0$) and not known at present.\\
\\
In the exponential band edge model (Eq.~\eqref{dos_exp}), we have
\begin{equation}\label{S2}
    S = \d{\varepsilon_c(B_z) - \varepsilon_f(B_z)}{eT}.
\end{equation}
Finally, in VRH, the thermopower depends on the structure of the
density of states at the Fermi level. For a linear density of
states, it is of the
form\cite{Zvyagin,Movaghar_Schirmacher,Wysokinski_Brenig}
\begin{equation}\label{S3}
    S \sim\d{k_B^2\rho(\varepsilon_f)(TT_0)^{\d{1}{2}}}{e}.
\end{equation}
The form which agrees with the experiments of Osinniy \emph{et
al}.\cite{Osinniy} is the band edge activated law Eq.~\eqref{S2}.
The activated form is also what is reported in a-Si:H, where the
exponential band edge model (Eq.~\eqref{dos_exp})
applies\cite{Dyalsingh_Kakalios}. The resistance measured by Osinniy
\emph{et al}.\cite{Osinniy} appears to also be activated. It seems
that we have a material class with band conduction, and where the
usual amorphous exponential density of states model applies. In this
material type, Eq.~\eqref{Emax} does not apply, and the integral
Eq.~\eqref{I1} is cut off at $k_BT$ above the mobility edge. The
transport is now controlled by the mobility edge, and Eq.~\eqref{S2}
follows. Magnetic thermopower measurements in the insulating regime
would be most useful, as they should obey Eqs.~\eqref{S2} and
~\eqref{S3} and help us understand the nature of the transport
mechanism. Pu \emph{et al}.\cite{Pu} have shown that in the metallic
regime, the magneto-thermopower is anisotropic (depends on the
orientation of $M_z$ to the current). This is due to spin-orbit
scattering. Note that the hopping thermopower, neglecting spin flip
energy transport, should also depend on the magnetic field via the
magnetic field dependence of the density of states $\rho$ and the
localization length $\alpha^{-1}$, with $T_0 =
24\times2.7\alpha^3(B_z)/[\pi k_B\rho(\varepsilon_f(B_z),B_z)]$.
Both spin-up and spin-down bands will contribute at the fermi level.
The localized states have a finite Hubbard $U$, so the full rigorous
treatment is not trivial, and needs focused attention as done in
Ref.~\onlinecite{Gruenewald2}, for the finite $U$ but B-field free
case.
\section{Band Region Above The Mobility Edge}\label{appen_3}
The mobility edge contribution to the conductivity, derived above,
is due to the fact that magnetic fields and other \emph{dephasing}
effects counteract the quantum Anderson localization process. The
B-field moves the mobility edge up for holes (down for electrons).
But in the delocalized region, electrons (holes) are also subject to
spin-disorder scattering which also causes strong magnetoresistance.
It may well be that the spin disorder band mechanism also
contributes when we have the insulating scenario, despite the high
activation energy. For completeness and rigor we therefore need to
also examine the transport dominated by carriers which are excited
below (above) the mobility edge in the real band region. When the
band region starts is not entirely clear but there is no reason why
spin disorder scattering should not start at the mobility edge as
well. The band transport should therefore include the spin disorder
scattering contribution as calculated using the Kubo
formula\cite{Van_Esch1}. We know that holes excited just below the
mobility edge are well described by the self-consistent CPA
solution\cite{Arsenault1,Arsenault2} so we can use it. The
contribution from the CPA band conductivity can be written
($\varepsilon_c = \varepsilon_c(B_z,M_z)$)
\begin{equation}\label{sig_CPAcontri_appen}
    \sigma_{xx} = \d{
    e^2}{\pi\Omega}\int_{-\infty}^{\varepsilon_c}d\varepsilon\left(-\d{\pd
    f(\varepsilon)}{\pd\varepsilon}\right)\rho(\varepsilon)D_{CPA}(\varepsilon,B_z),
\end{equation}
where
\begin{equation}
    \rho(\varepsilon)D_{CPA}(\varepsilon) =
    \d{\hbar}{N}\sum_{s}\int d\varepsilon_kv_x^2\left[ \text{Im}\{ G_{k,s}(\varepsilon) \}
    \right]^2\delta(\varepsilon-\varepsilon_{k,s}).
\end{equation}
Also we have
\begin{equation}
    G_{k,s}(\varepsilon) = \d{1}{\varepsilon-\varepsilon_{k,s}-\Sigma_{s,CPA}(\varepsilon)}
\end{equation}
In Fig.~\ref{fig:MR_B_J035x053EM0} and
Fig.~\ref{fig:resis_B0x053EM0_B_T}, we show the calculated curves
for a parameter range which would correspond to transport at the
mobility edge. The resistance results shown, multiplied by the
activation factor
$\text{exp}\left[\d{\varepsilon_f-\varepsilon_c}{k_BT}\right]$,
would give us a measure  of the spin disorder band contribution to
the magnetoresistance at low temperatures. We note that the extended
state spin-magnetoresistance is negative almost everywhere. This is
a result of the fully self-consistent single-site approximation
which treats both the magnetism and the transport on the same
footing. The intuitive reason is that the material is more ordered
when the local moments are aligned. When disordered, the holes are
scattered from many types of potentials corresponding to the 6
possible $S_z$ orientation of the local Mn 5/2 moment. When a
magnetic field is applied, it enhances the order, and this reduces
the resistance. Van Esch \emph{et al}.\cite{Van_Esch1} have shown
that the resistance decreases with magnetization $M_z(B_z,T)$. We
have used their formula to analytically simulate the CPA result
coming from spin disorder scattering in Eq.~\eqref{pd_scatt}.

\end{document}